\documentclass{llncs}
\usepackage[utf8]{inputenc}

\usepackage{listings,color}

\usepackage{pgf}
\usepackage{paralist}
\usepackage{amsmath}
\usepackage{multirow}

\newcommand{\setjavamode}{\lstset{language=Java,basicstyle={\tt\small},keywordstyle=\bfseries}}
\newcommand{\setcamlmode}{\lstset{language=Caml,basicstyle={\tt\small},keywordstyle=\bfseries}}

\newcommand{\remarques}[1]{}
\newcommand{\sawja}{\textsc{Sawja}}
\newcommand{\javalib}{\textsc{Javalib}}
\newcommand{\Soot}{\textsc{Soot}}
\newcommand{\Wala}{\textsc{Wala}}
\newcommand{\OCaml}{\textsc{OCaml}}
\newcommand{\Coq}{\textsc{Coq}}
\newcommand{\Java}{\textsc{Java}}
\renewcommand{\texttt}[1]{\lstinline!#1!}

\newcommand{\afterfigure}{\vspace*{-1em}}
\newcommand{\aftertable}{\vspace*{-3em}}
\newcommand{\beforecaption}{\vspace*{-1em}}

\begin{document}

\title{\sawja{}: Static Analysis Workshop for \Java{}}

\author{Laurent Hubert\inst{1} \and Nicolas Barré\inst{2} \and
  Frédéric Besson\inst{2} \and Delphine Demange\inst{3} \and Thomas
  Jensen\inst{2} \and Vincent Monfort\inst{2} \and David
  Pichardie\inst{2} \and Tiphaine Turpin\inst{2}}

\institute{CNRS/IRISA, France \and INRIA Rennes - Bretagne Atlantique,
  France \and ENS Cachan - Antenne de Bretagne/IRISA, France}

\maketitle{}

\setjavamode

\begin{abstract}
  Static analysis is a powerful technique for automatic verification
  of programs but raises major engineering challenges when developing
  a full-fledged analyzer for a realistic language such as \Java{}.
  Efficiency and precision of such a tool rely partly on low level
  components which only depend on the syntactic structure of the
  language and therefore should not be redesigned for each
  implementation of a new static analysis.  This paper describes the
  \sawja{} library: a static analysis framework fully compliant with
  \Java{} 6 which provides \OCaml{} modules for efficiently
  manipulating \Java{} bytecode programs. We present the main features
  of the library, including i) efficient functional data-structures
  for representing program with implicit sharing and lazy parsing, ii)
  an intermediate stack-less representation, and iii) fast computation
  and manipulation of complete programs. We provide experimental
  evaluations of the different features with respect to time, memory
  and precision.
\end{abstract}


\section*{Introduction}
\label{sec:introduction}

Static analysis is a powerful technique that enables automatic
verification of programs with respect to various properties such as
type safety or resource consumption. One particular well-known example
of static analysis is given by the \Java{} Bytecode Verifier (BCV),
which verifies at loading time that a given \Java{} class (in bytecode
form) is type safe.  Developing an analysis for a realistic
language such as \Java{} is a major engineering task, challenging both
the
companies that want to build robust commercial tools and the research
scientists who want to quickly develop prototypes for demonstrating
new ideas.
The efficiency and the precision of any static analysis depend on the
low-level components which manipulates the class hierarchy, the call
graph, the intermediate representation (IR), etc. These components are
not specific to one particular analysis, but they are far too often
re-implemented in an \emph{ad hoc} fashion, resulting in analyzers
whose overall behaviour is sub-optimal (in terms of efficiency or
precision). We argue that it is an integral part of automated software
verification to address the issue of how to program a static analysis
platform that is at the same time efficient, precise and generic, and
that can facilitate the subsequent implementation of specific
analyzers.

This paper describes the \sawja{} library ---~and its sub-component
\javalib{}~--- which provides \OCaml{} modules for efficiently
manipulating \Java{} bytecode programs. The library is developed under
the GNU Lesser General Public License and is freely available at
  \url{http://sawja.inria.fr/}.

\sawja{} is implemented in \OCaml~\cite{ocaml}, a  strongly
typed functional language whose automatic memory
management (garbage collector), strong typing and pattern-matching facilities
 make particularly well suited for implementing program processing
 tools. In particular, it has been successfully
used for programming compilers (e.g.,  Esterel~\cite{Esterel})
and static analyzers (e.g.,  Astrée~\cite{Astree}).

The main contribution of the \sawja{} library is to provide, in a
unified framework, several features that allow rapid prototyping of
efficient static analyses while handling all the subtleties of the
\Java{} Virtual Machine (JVM) specification~\cite{lindholm99:jvm_spec}.
The main features of \sawja{} are:
\begin{itemize}
\item parsing of \texttt{.class} files into \OCaml{} structures and
  unparsing of those structures back into \texttt{.class} files;
\item decompilation of the bytecode into a high-level stack-less IR;
\item sharing of complex objects both for memory saving and efficiency
  purpose (structural equality becomes equivalent to pointer equality
  and indexation allows fast access to tables indexed by class, field
  or method signatures, etc.);
\item the determination of the set of classes constituting a complete
  program (using several algorithms, including Rapid Type Analysis
  (RTA)~\cite{BaconS96});
\item a careful translation of many common definitions of the JVM
  specification, e.g., about the class hierarchy, field and
  method resolution and look-up, and intra- and inter-procedural
  control flow graphs.
\end{itemize}

This paper describes the main features of \sawja{} and their
experimental evaluation. Sect.~\ref{sec:related-work-1} gives an
overview of existing libraries for manipulating \Java{} bytecode.
Sect.~\ref{sec:low-level} describes the representation of classes,
Sect.~\ref{sec:ir} presents the intermediate representation of
\sawja{} and Sect.~\ref{sec:complete-program} presents the parsing of
complete programs.

\section{Existing Libraries for Manipulating \Java{} Bytecode}
\label{sec:related-work-1}


Several similar libraries have already been developed so far and some of them
provide features similar to some of \sawja{}'s. All of them, except
\textsc{Barista}, are written in \Java{}.

The Byte Code Engineering
Library\footnote{\url{http://jakarta.apache.org/bcel/}}(\textsc{BCEL})
and \textsc{ASM}\footnote{\url{http://asm.ow2.org/}} are open source
\Java{} libraries for generating, transforming and analysing \Java{}
bytecode classes. These libraries can be used to manipulate classes at
compile-time but also at run-time, e.g., for dynamic class
generation and transformation.  \textsc{ASM} is particularly optimised
for this latter case: it provides a visitor pattern which makes
possible local class transformations without even building an
intermediate parse-tree.  Those libraries are well adapted to
instrument \Java{} classes but lack important features essential for
the design of static analyses.  For instance, unlike \sawja{}, neither
\textsc{BCEL} nor \textsc{ASM} propose a high-level intermediate
representation (IR) of bytecode instructions. Moreover, there is no
support for building the class hierarchy and analysing complete
programs.  The data structures of \javalib{} and \sawja{} are also
optimized to manipulate large programs.

The \textsc{Jalapeño} Optimizing Compiler~\cite{jalapeno99} which is
now part of the \textsc{Jikes} RVM relies on two IR (low and
high-level IR) in order to optimize bytecode. The high-level IR is a
3-address code. It is generated using a symbolic evaluation technique
described in~\cite{whaley99}. The algorithm we use to generate our IR
is similar. Our algorithm works on a fixed number of passes on the
bytecode while their algorithm is iterative. The \textsc{Jalapeño}
high-level IR language provides explicit check instructions for common
run-time exceptions (e.g., \texttt{null\_check},
\texttt{bound\_check}), so that they can be easily moved or eliminated
by optimizations. We use similar explicit checks but to another end:
static analyses definitely benefit from them as they ensure
expressions are error-free.

\Soot{}~\cite{soot99} is a \Java{} bytecode optimization framework
providing three IR: Baf, Jimple and Grimp.  Optimizing \Java{}
bytecode consists in successively translating bytecode into Baf,
Jimple, and Grimp, and then back to bytecode, while performing diverse
optimizations on each IR.  Baf is a fully typed, stack-based
language. Jimple is a typed stack-less 3-address code and Grimp is a
stack-less representation with tree expressions, obtained by
collapsing Jimple instructions.
The IR in \sawja{} and \Soot{} are very similar but are obtained by
different transformation techniques. They are experimentally compared
in Sect.~\ref{sec:ir}.  Several state-of-the-art control-flow
analyses, based on points-to analyses, are available in Soot through
Spark~\cite{lhot.hend03} and Paddle~\cite{LhotakHendren:evalBDD}. Such
libraries represent a coding effort of several man-years. To this
respect, \sawja{} is less mature and only proposes simple (but
efficient) control-flow analyses.

\Wala{}~\cite{wala} is a \Java{} library dedicated to static analysis
of \Java{} bytecode. The framework is very complete and provides
several modules like control flow analyses, slicing analyses, an
inter-procedural dataflow solver and a IR in SSA form. \Wala{} also
includes a front-end for other languages like \Java{} source and
\textsc{JavaScript}. \Wala{} and its IBM predecessor DOMO have been
widely used in research prototypes. It is the product of the long
experience of IBM in the area. Compared to it, \sawja{} is a more
recent library with less components, especially in terms of static
analyses examples.  Nevertheless, the results presented in
Sect.~\ref{sec:complete-program} show that \sawja{} loads programs
faster and uses less memory than \Wala{}.  For the moment, no SSA IR
is available in \sawja{} but this could easily be added.

\textsc{Julia}~\cite{Spoto05} is a generic static analysis tool for
\Java{} bytecode based on the theory of abstract interpretation. It
favors a particular style of static analysis specified with respect to
a denotational fixpoint semantics of \Java{} bytecode.  Initially free
software, \textsc{Julia} is not available anymore. 

\textsc{Barista}~\cite{barista} is an \OCaml{} library used in the
\textsc{OCaml-Java} project. It is designed to load, construct,
manipulate and save \Java{} class files. \textsc{Barista} also
features a \Java{} API to access the library directly from
\Java{}. There are two representations: a low-level representation,
structurally equivalent to the class file format as defined by Sun,
and a higher level representation in which the constant pool indices
are replaced by the actual data and the flags are replaced by
enumerated types.  Both representations are less factorized than in
\javalib{}
and, unlike \javalib{}, \textsc{Barista} does not encode the
structural constraints into the \OCaml{} structures.  Moreover, it is
mainly designed to manipulate single classes and does not offer the
optimizations required to manipulate sets of classes (lazy parsing,
hash-consing, etc).


%

\section{High-level Representation of Classes}
\label{sec:low-level}

\sawja{} is built on top of \javalib{}, a \Java{} bytecode parser
providing basic services for manipulating class files, i.e., an
optimised high-level representation of class files, pretty printing
and unparsing of class files.\footnote{\javalib{} is a sub-component
  of \sawja{}, which, despite being tightly integrated in \sawja{},
  can also be used as an independent library.  It was initiated by
  Nicolas Cannasse before 2004 but, since 2007, we have largely
  extended the library.  We are the current maintainers of the
  library.}  \javalib{} handles all aspects of class files, including
stackmaps (J2ME and \Java{}~6) and \Java{}~5 annotation attributes. It
is made of three modules:
\fbox{\href{http://javalib.gforge.inria.fr/doc/javalib-api/Javalib.html}{Javalib}},
\fbox{\href{http://javalib.gforge.inria.fr/doc/javalib-api/JBasics.html}{JBasics}},
and
\fbox{\href{http://javalib.gforge.inria.fr/doc/javalib-api/JCode.html}{JCode}}\footnote{In
  the following, we use boxes around \javalib{} and \sawja{} module
  names to make clickable links to the on-line API documentation}.

Representing class files constitutes the low-level part of a bytecode
manipulation library. 
Our design choices are  driven by a set of
principles which are explained below.  


\paragraph{Strong typing}
We use the OCaml type system to explicit as much as possible the
structural constraints of the class file format. For example, interfaces are only
signaled by a flag in the \Java{} class file format and this requires to check
several consistency constraints between this flag and the content of
the class (interface methods must be abstract, the super-class must be
\lstinline{java.lang.Object}, etc.).  Our representation distinguishes
classes and interfaces and these constraints are therefore expressed
and enforced at the type level. This has two advantages. First, this
lets the user concentrate on admissible class files, by reducing the
burden of handling illegal cases. Second, for the generation (or
transformation) of class files, this provides good support for
creating correct class files.

\begin{figure}[t]\centering
\pgfimage[width=.8\linewidth]{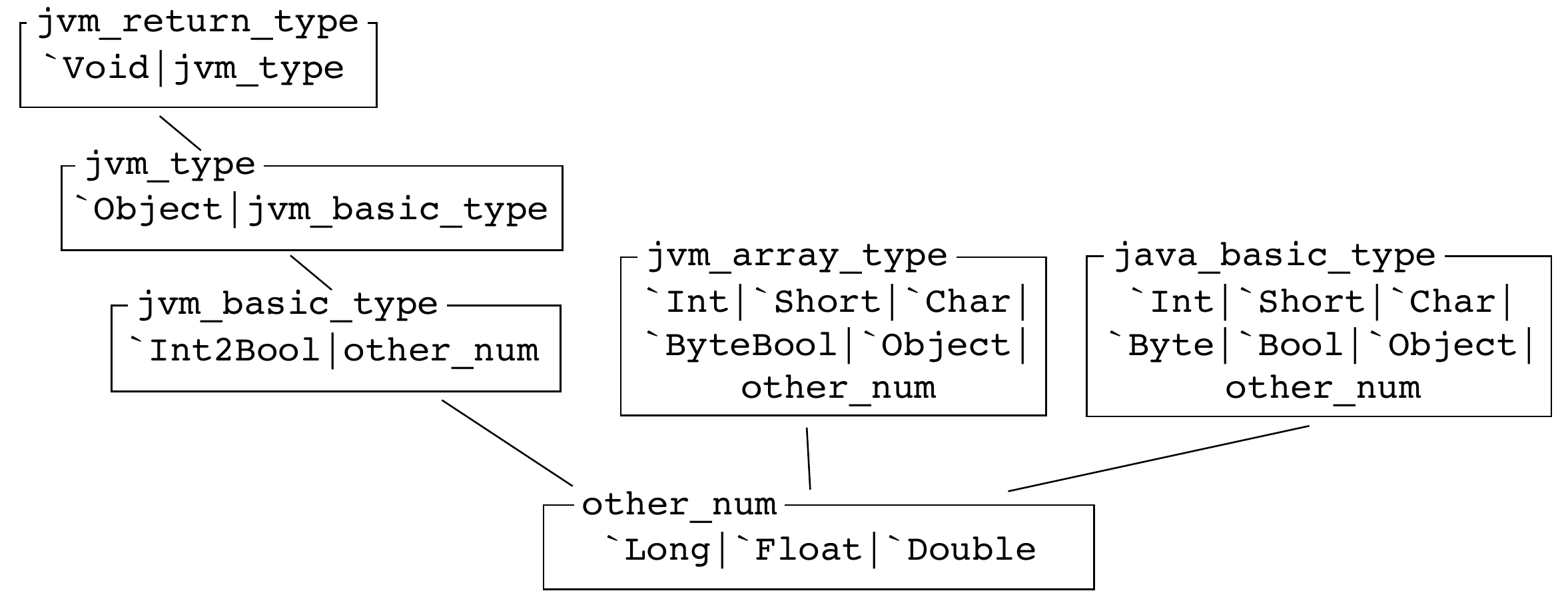}

\beforecaption{}\caption{Hierarchy of \Java{} bytecode types. 
Links represent the subtyping
relation enforced by polymorphic variants (for example, the type
\lstinline!jvm_type! is defined by 
\lstinline!type jvm_type = [ |`Object |jvm_basic_type ]!).}
\label{fig:polyvariant}
\afterfigure{}
\end{figure}

\paragraph{Factorization}\setcamlmode 
Strong typing sometimes lacks flexibility and can lead to unwanted
code duplication.
An example is the use of several, distinct notions of types in class
files at different places (JVM types, \Java{} types, and JVM array
types). We factorize common elements as much as possible, sometimes by
a compromise on strong typing, and by relying on specific language
features such as polymorphic variants\footnote{Polymorphic variants
  are a particular notion of enumeration which allows the sharing of
  constructors between types.}. Fig.~\ref{fig:polyvariant} describes
the hierarchy formed by these types. This factorization principle
applies in particular to the representation of op-codes: many
instructions exist whose name only differ in the JVM type of their
operand, and variants exist for particular immediate values
(e.g., \texttt{iload}, \texttt{aload}, \texttt{aload\_$n$},
etc.).  In our representation they are grouped into families with the
type given as a parameter (\lstinline!OpLoad of jvm_type * int!).
\setjavamode
\paragraph{Lazy Parsing}

To minimise the memory footprint, method bodies are parsed on demand
when their code is first accessed.  This is almost transparent to the
user thanks to the \lstinline!Lazy! \OCaml{} library but is important
when dealing with very large programs. It follows that dead code (or
method bodies not needed for a particular analysis) does not cause any
time or space penalty.

\paragraph{Hash-consing of the Constant Pool}
For a \Java{} class file, the constant pool is a table which gathers
all sorts of data elements appearing in the class, such as Unicode
strings, field and method signatures, and primitive values. Using the
constant pool indices instead of actual data reduces the class files
size.  This low-level aspect is abstracted away by the \javalib{}
library, but the sharing is retained and actually strengthened by the
use of \emph{hash-consing}.
Hash-consing~\cite{ershov58:hash_consing} is a general technique for
ensuring maximal sharing of data-structures by storing all data in a
hash table.  It ensures unicity in memory of each piece
of data and allows to replace structural equality tests by tests on
pointers.  In \javalib{}, it is used for constant pool items that are
likely to occur in several class files, i.e., class names, and
field and method signatures. Hash-consing is global: a class name like
\lstinline{java.lang.Object} is therefore shared across all the parsed
class files.
For \javalib{}, our experience shows that hash-consing is always a
winning strategy; it reduces the memory footprint and is almost
unnoticeable in terms of running time\footnote{The indexing time is
  compensated by a reduced stress on the garbage collector.}. We
implement a variant which assigns hash-consed values a unique
(integer) identifier.  It enables optimised algorithms and
data-structures.  In particular, the \javalib{} API
features sets and maps of hash-consed values based on Patricia
trees~\cite{morrison68:patricia_trees}, which are a type of prefix
tree.  Patricia trees are an efficient purely functional
data-structure for representing sets and maps of integers,
i.e., identifiers of hash-consed values. They exhibit good
sharing properties that make them very space efficient.  Patricia
trees have been proved very efficient for implementing flow-sensitive
static analyses where sharing between different maps at different
program points is crucial.
On a very small benchmark computing the transitive closure of a call
graph, the indexing makes the computation time four times smaller.
Similar data-structures have been used with success in the Astrée
analyzer~\cite{Astree}.

\paragraph{Visualization}
\sawja{} includes functions to print the content of a class into
different formats.  A first one is simply raw text, very close to the
bytecode format as output by the \texttt{javap} command (provided with
Sun's JDK).

A second format is compatible with Jasmin~\cite{jvm}, a \Java{}
bytecode assembler. This format can be used to generate incorrect
class files (e.g., during a \Java{} virtual machine testing),
which are difficult to generate with our framework. The idea is then,
using a simple text editor, to manually modify the Jasmin files output
by \sawja{} and then to assemble them with Jasmin, which does not
check classes for structural constraints.

Finally, \sawja{} provides an HTML output. It allows displaying class
files where the method code can be folded and unfolded simply by
clicking next to the method name. It also makes it possible to open
the declaration of a method by clicking on its signature in a method
call, and to know which method a method overrides, or by which methods
a method is overridden, etc.  User information can also be displayed
along with the code, such as the result of a static analysis.  From
our experience, it allows a faster debugging of static analyses.

\section{Intermediate Representation}
\label{sec:ir}
\remarques{add what is brought by the theorem? }
The JVM is a stack-based virtual machine and the intensive use of the
operand stack makes it difficult to adapt standard static analysis
techniques that have been first designed for more classic
variable-based codes. Hence, several bytecode optimization and
analysis tools work on a bytecode \emph{intermediate representation}
(IR) that makes analyses simpler~\cite{jalapeno99,soot99}.
Surprisingly, the semantic foundations of these 
transformations have received little attention. The transformation
that is informally presented here has been formally studied and proved
 semantics-preserving in~\cite{DemangeJP09}.

\subsection{Overview of the IR Language}

\begin{figure}[t]\centering
\pgfimage[width=\linewidth]{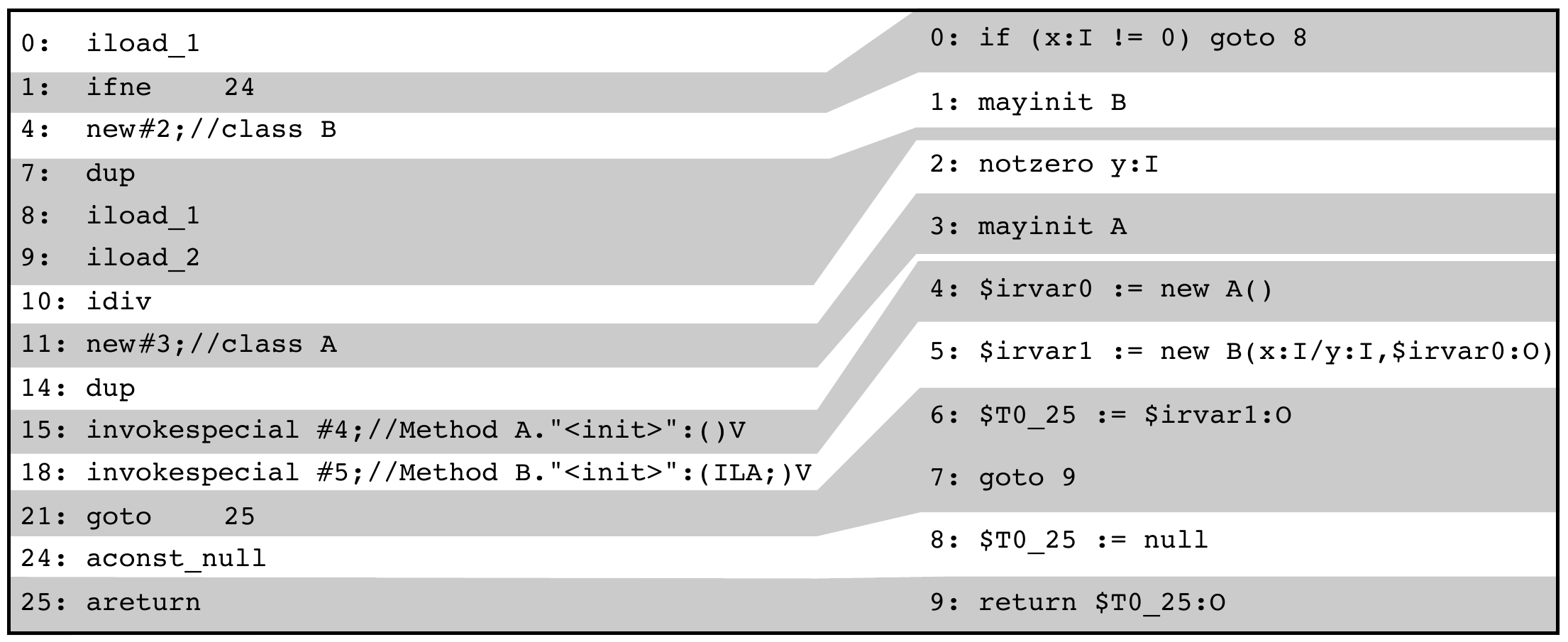}
\beforecaption{}\caption{Example of bytecode (left) (obtained with \lstinline{javap -c}) and its corresponding IR (right). Colors make explicit the boundaries of related code fragments.  }
\label{fig:example}
\afterfigure
\end{figure}

Fig.~\ref{fig:example} gives the bytecode and IR versions of the simple method
\begin{lstlisting}
B f(int x, int y) { return (x==0)?(new B(x/y, new A())):null;}   
\end{lstlisting}

The bytecode version reads as follows : the value of the first
argument \lstinline{x} is pushed on the stack at program point $0$. At
point $1$, depending on whether \lstinline{x} is zero or not, the
control flow jumps to point $4$ or $24$ (in which case the value
\lstinline{null} is returned).  At point $4$, a new object of class
\lstinline{B} is allocated in the heap and its reference is pushed on
top of the operand stack. Its address is then duplicated on the stack
at point $7$. Note the object is \emph{not initialized} yet. Before
the constructor of class \lstinline{B} is called (at point $18$), its
arguments must be computed: lines $8$ to $10$ compute the division of
\lstinline{x} by \lstinline{y}, lines $11$ to $15$ construct an object
of class \lstinline{A}. At point $18$, the non-virtual method
\lstinline{B} is called, consuming the three top elements of the
stack. The remaining reference of the \lstinline{B} object is left on
the top of the stack and represents from now on an \emph{initialized}
object.

The right side of Fig.~\ref{fig:example} illustrates the main features
of the IR language.~\footnote{For a complete description of the IR
  language syntax, please refer to the API documentation of the
  \fbox{\href{http://javalib.gforge.inria.fr/doc/sawja-api/JBir.html}{JBir}}
  module.  A 3-address representation called
  \fbox{\href{http://javalib.gforge.inria.fr/doc/sawja-api/A3Bir.html}{A3Bir}}
  is also available where each expression is of height at most~$1$.}
First, it is \emph{stack-less} and manipulates \emph{structured
  expressions}, where variables are annotated with \emph{types}. For
instance, at point $0$, the branching instruction contains the
expression \lstinline{x:I}, where \lstinline{I} denotes the type of
\Java{} integers. Another example of recovered structured expression
is \lstinline{x:I/y:I} (at point $5$). Second, expressions are
\emph{error-free} thanks to explicit checks: the
instruction \lstinline{notzero y:I} at point $2$ ensures that
evaluating \lstinline{x:I/y:I} will not raise any error. Explicit
checks additionally guarantee that the order in which
\emph{exceptions} are raised in the bytecode is preserved in the IR.
Next, the object creation process is syntactically simpler in the IR
because the two distinct phases of
\begin{inparaenum}[(i)]
\item allocation and
\item constructor call
\end{inparaenum}
are merged by \emph{folding} them into a single IR instruction (see
point $4$). In order to simplify the design of static analyses on the
IR, we forbid side-effects in expressions. Hence, the nested
object creation at source level is decomposed into two 
assignments (\lstinline!$irvar0! and \lstinline!$irvar1! are temporary
variables introduced by the transformation). Notice that because of
side-effect free expressions, the order in which the \lstinline!A! and
\lstinline!B! objects are allocated must be reversed. Still, the IR
code is able to preserve the \emph{class initialization order} using
the dedicated instruction \lstinline!mayinit! that calls the static
class initializer whenever it is required.

\subsection{IR Generation}
\remarques{Reviewer:Similarly, Section 3.2 is difficult to follow and
  does not bring anything very useful into the whole picture.}%
The purpose of the \sawja{} library is not only static analysis but
also lightweight verification~\cite{Rose:lightweight}, i.e.,
the verification of the result of a static analysis in a single pass
over the method code.  To this end, our transforming algorithm
operates in a fixed number of passes on the bytecode, i.e.,
without performing fixpoint iteration.

\Java{} subroutines (bytecodes \texttt{jsr}/\texttt{ret}) are inlined.
Subroutines have been pointed out by the research community as raising
major static analysis difficulties~\cite{StataA98}.  Our restricted
inlining algorithm cannot handle nested subroutines but is
sufficient to inline all subroutines from Sun's \Java{}~7 JRE.

The IR generation is based on a symbolic execution of the bytecode:
each bytecode modifies a stack of symbolic expressions, and
potentially gives rise to the generation of IR instructions. For
instance, bytecodes at lines $8$ and $9$ (left part of
Fig.~\ref{fig:example}) respectively push the expressions
\lstinline!x! and \lstinline!y! on the symbolic stack (and do not
generate IR instructions). At point $10$, both expressions are
consumed to build both the IR explicit check instruction and the
expression \lstinline!x/y! which is then pushed, as a result, on the
symbolic stack.  The non-iterative nature of our algorithm makes the
transformation of jumping instructions non-trivial. Indeed, during the
transformation, the output symbolic stack of a given bytecode is used
as the entry symbolic stack of all its successors. At a join point, we
thus must ensure that the entry symbolic stack is the same regardless
of its predecessors. The idea is here to empty the stack at branching
points and restore it at join points. More details can be found
in~\cite{DemangeJP09}.
  IR expression types are computed using a standard type
inference algorithm similar to what is done by the BCV. It only
differs in the type domain we used, which is less precise, but does
not require iterating. This additionally allows us interleaving
expression typing with the IR generation, thus resulting in a gain in
efficiency. This lack of precision could be easily filled in using the
stackmaps proposed in the \Java{}~6 specification.



\subsection{Experiments}

We validate the \sawja{} IR with respect to two criteria. We first
evaluate the time efficiency of the IR generation from \Java{}
bytecode. Then, we show that the generated code contains a reasonable
number of local variables.
We additionally compare our tool with the \Soot{} framework.  Our
benchmark libraries are real-size \Java{} code available in
\texttt{.jar} format. This includes Javacc $4.0$ (\Java{} Compiler
Compiler), JScience $4.3$ (a comprehensive \Java{} library for the
scientific community), the \Java{} runtime library $1.5.0\_12$ and \Soot{}
 $2.2.3$.

\subsubsection{IR Generation Time}
\begin{figure}[t]
  \begin{minipage}[t]{.5\textwidth}
  \centering
  \pgfimage[width=\textwidth]{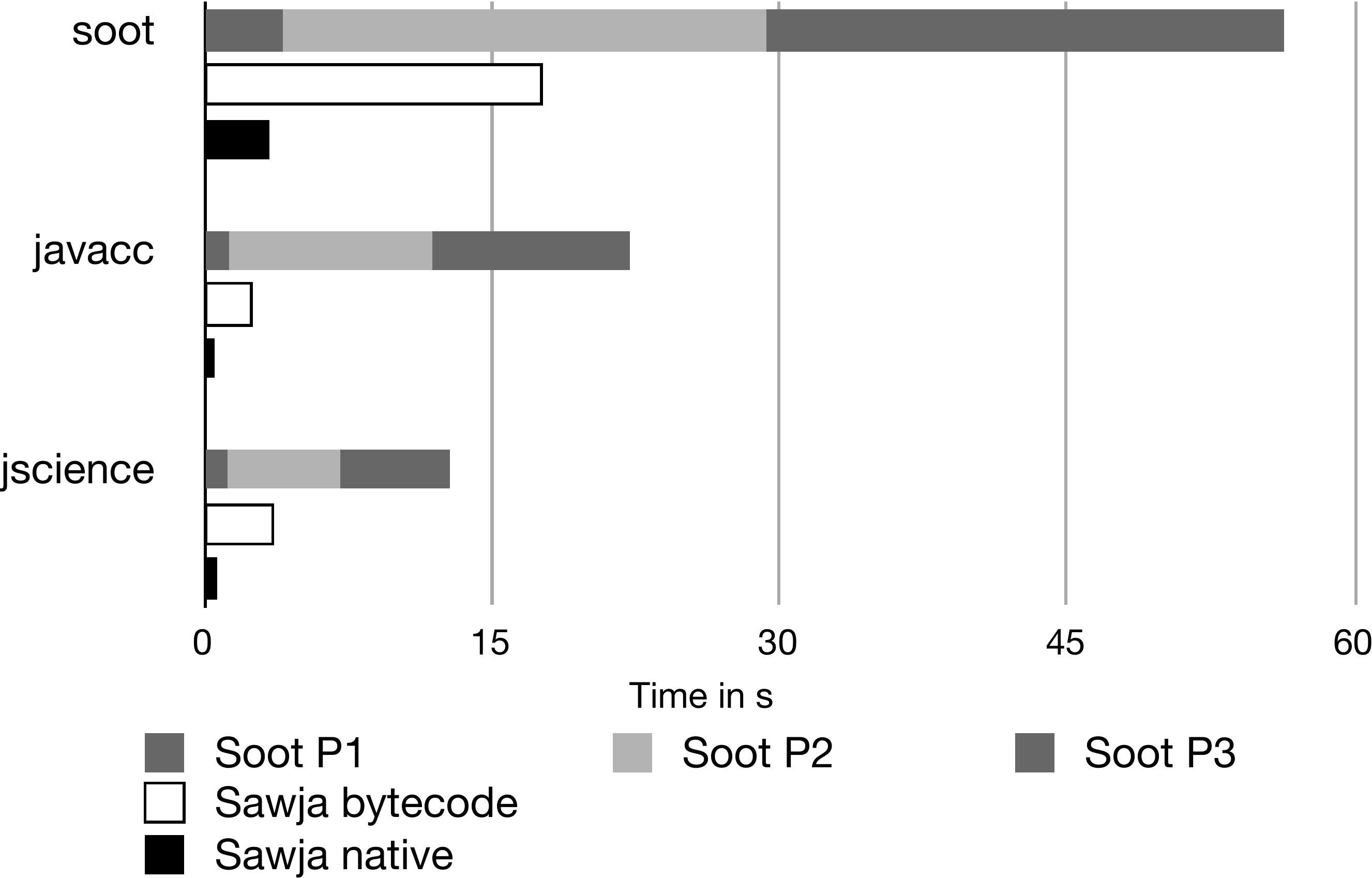}
  \beforecaption{}\caption{\sawja{} and \Soot{} IR generation times}
  \label{fig:efficiency}
  \end{minipage}
  \hfill
  \begin{minipage}[t]{0.45\textwidth}
    \centering
    \pgfimage[width=\textwidth]{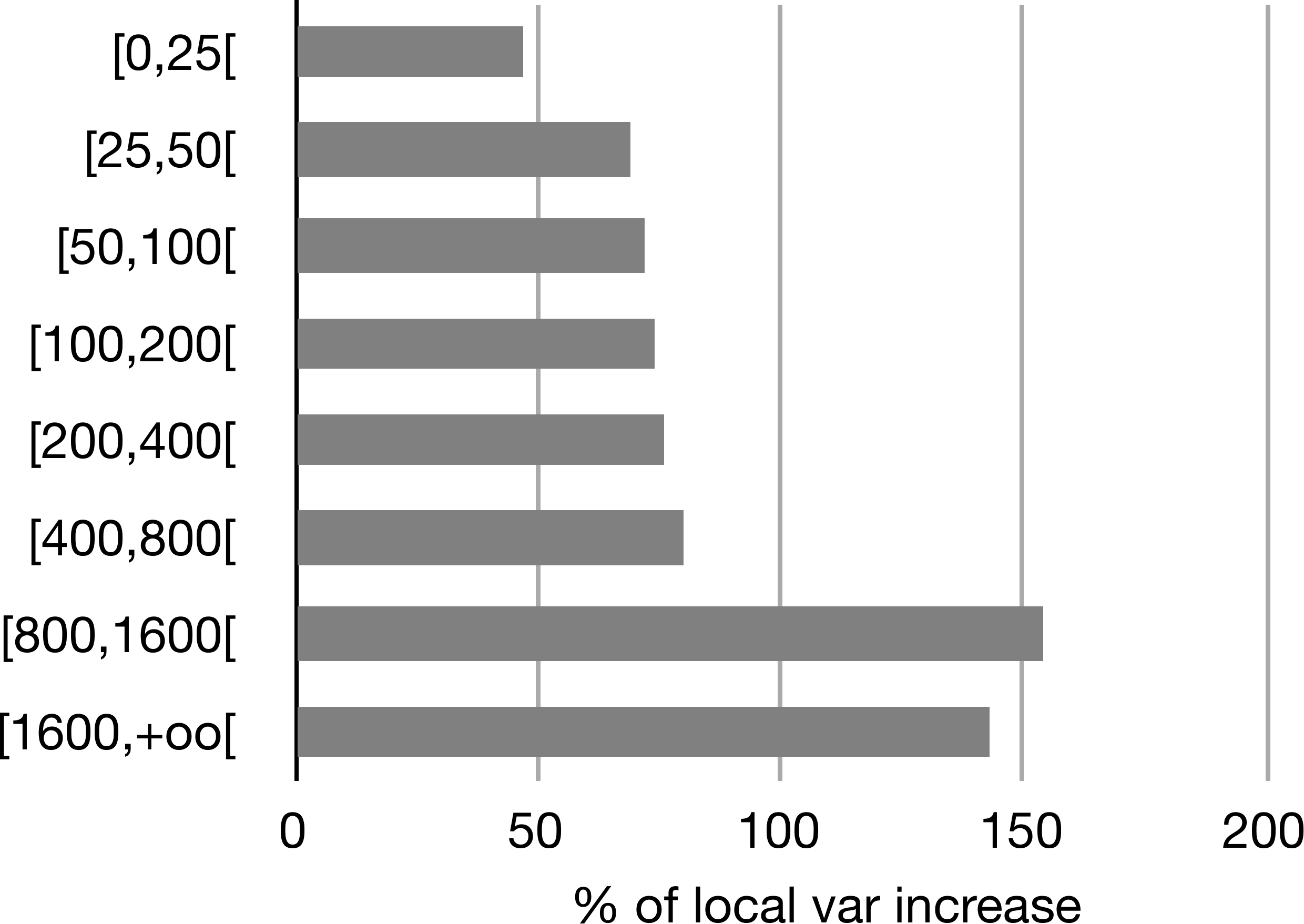}
    \beforecaption{}\caption{\sawja{}: local variable increase}
    \label{fig:compact1}
  \end{minipage}
\afterfigure
\end{figure}

In order to be usable for lightweight verification, the bytecode
transformation must be efficient. This is mainly why we avoid
iterative techniques in our algorithm.  We compare the transformation
time of our tool with the one of \Soot{}.  The results are given in
Fig.~\ref{fig:efficiency}.  For each benchmark library\footnote{For
  scale reason, the \Java{} runtime library measures are not shown
  here.}, we compare our running time for transforming all classes
with the running time of \Soot. Here, we choose to generate with
\Soot{} the Grimp representation of classes\footnote{The \Soot{}
  transformation is without any optimisation option.}, the closest IR
to ours that \Soot{} provides. Grimp allows expressions with
side-effects, hence expressions are somewhat more aggregated than in
our IR.  However, this does not inverse the trend of results.  We rely
on the time measures provided by \Soot{}, from which we only keep
three phases: generation of naive Jimple 3-address code (P1), local
def/use analysis used to simplify this naive code (P2), and
aggregation of expressions to build Grimp syntax (P3).  (Other phases,
like typing, are not directly relevant.)
Unlike \Java{} code, \OCaml{} code is usually executed in native form.
For the comparison not to be biaised, we compare execution times of
both tools in bytecode form and also give the execution time of
\sawja{} in native form.
These experiments show that \sawja{} (both in bytecode and native
mode) is very competitive with respect to \Soot{}, in terms of
computation efficiency. This is mainly due to the fact that, contrary
to \Soot{}, our algorithm is non-iterative.

\subsubsection{Compactness of the Obtained Code}
Intermediate representations rely on temporary variables in order to
remove the use of operand stack and generate side-effect free
expressions. The major risk here is an explosion in the number of new
variables when transforming large programs.


In practice our tool stays below doubling the number of local
variables, except for very large methods ($> 800$ bytecodes). 
Fig.~\ref{fig:compact1} presents the percentage of local
variable increase induced by our transformation, for each method of our
 benchmarks, and sorting results according to the method size (indicated by numbers in brackets). 
The number of new variables stays manageable and 
we believe it could be further reduced using standard optimization techniques,
as those employed by \Soot{}, but this would require to iterate on
each method.

\begin{figure}[t]
  \centering
  \pgfimage[width=\textwidth]{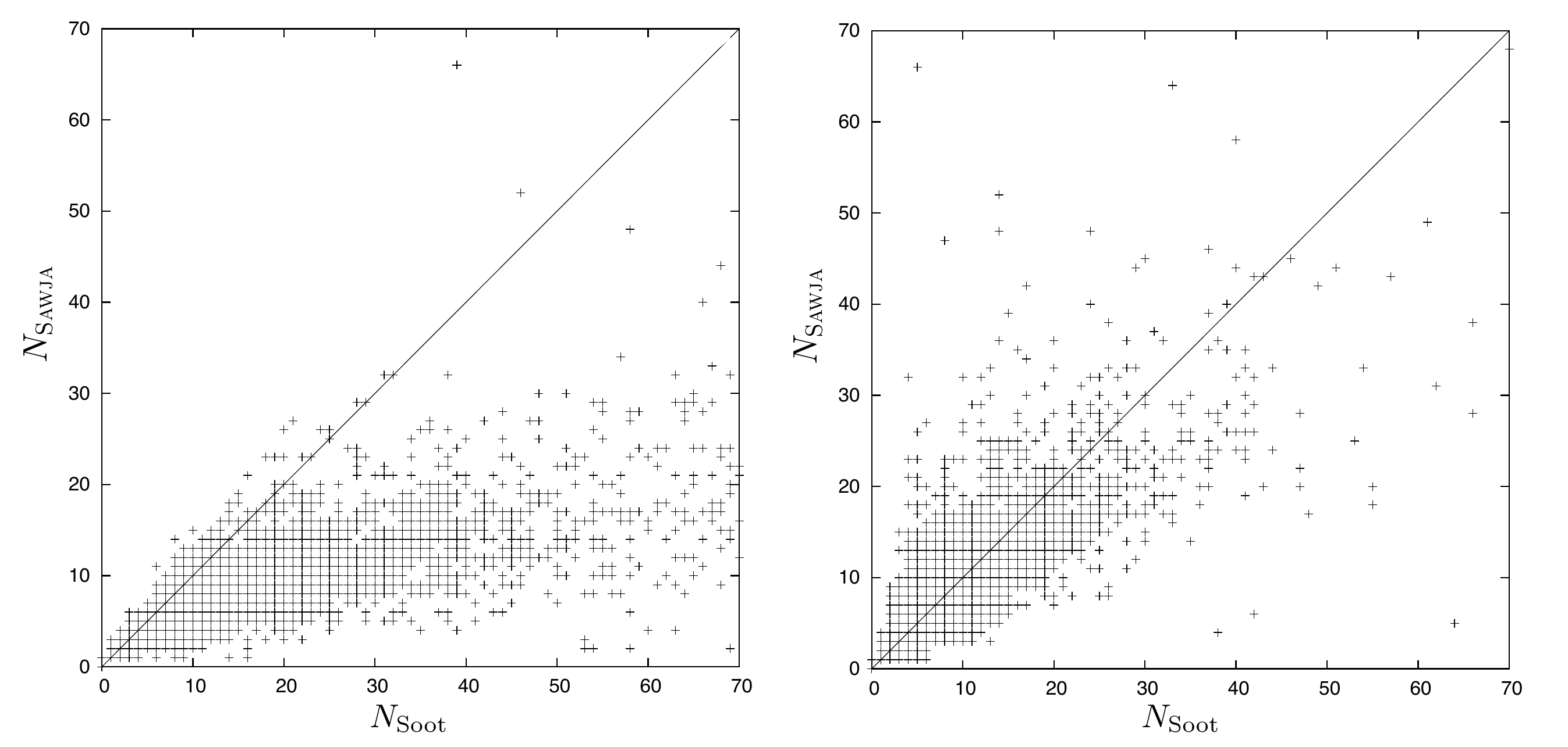}
  \beforecaption{}\caption{Local variable increase ratio between \sawja{} and \Soot.}
  \label{fig:compact2}
\afterfigure
\end{figure}

We have made a direct comparison with the \Soot{} in terms of the
local variable increase.  Fig.~\ref{fig:compact2} presents two
measures. For each method of our benchmarks we count the number
$N_{\it{\sawja{}}}$ of local variables in our IR code and the number
$N_{\it{Soot}}$ of local variables in the code generated by \Soot{}. A
direct comparison of our IR against Grimp code is difficult because it
allows expressions with side-effects, thus reducing the amount of
required variables.  Hence, in this experiment, the comparison is made
between \Soot{}'s 3-address IR (Jimple) and our 3-address IR. For each
method we draw a point of coordinate $(N_{\it{Soot}},N_{\it{\sawja}})$
and see how the points are spread out around the first bisector. For
the left diagram, \Soot{} has been launched with default options. For
the right diagram, we added to the \Soot{} transformation the local
packer (\texttt{-p jb.lp enabled:true} \Soot{} option) that
reallocates local variables using use/def information.  Our
transformation competes well, even when \Soot{} uses this last
optimization. We could probably improve this ratio using a similar
packing.

\section{Complete Programs}
\label{sec:complete-program}



Whole program analyses require a model of the global control-flow
graph of an entire Java{} program. For those, \sawja{} proposes
the notion of \emph{complete programs}. Complete programs are equipped
with a high-level API for navigating the control-flow graph and are
constructed by a preliminary control-flow analysis.


\subsection{API of Complete Programs}
\label{sec:api-complete-prog}
\sawja{} represents a complete program by a record.  The field
\lstinline{classes} maps a class name to a class node in the class
hierarchy.  The class hierarchy is such that any class referenced in
the program is present.  The field \lstinline{parsed_methods} maps a
fully qualified method name to the class node declaring the method and
the implementation of the method.  The field
\lstinline{static_lookup_method} returns the set of target methods of
a given field.  As it is computed statically, the target methods are
an over-approximation.

The API allows navigating the intra-procedural graph of a method
taking into account jumps, conditionals and exceptions.  Although
conceptually simple, field and method resolution and the different
method look-up algorithms (corresponding to the instructions
\lstinline!invokespecial!, \lstinline!invokestatic!,
\lstinline!invokevirtual!, \lstinline!invokeinterface!)  are critical
for the soundness of inter-procedural static analyses.  In \sawja{},
great care has been taken to ensure an implementation fully compliant
with the JVM specification.


\subsection{Construction of Complete Programs}

Computing the exact control-flow graph of a \Java{} application is
undecidable and computing a precise (over-)approximation of it 
is still computationally challenging. It is a field of active research
(see for instance~\cite{LhotakHendren:evalBDD,DOOP}).
A complete program is computed by:
\begin{inparaenum}[(1)]
\item initializing the set of reachable code to the entry points of the
  program,
\item computing the new call graph, and
\item if a (new) edge of the call graph points to a new node, adding
  the node to the set of reachable code and repeating from step (2).
\end{inparaenum}
The set of code obtained when this iteration stops is an
over-approximation of the complete program.

Computing the call graph is done by resolving all reachable method
calls.  Here, we use the functions provided in the \sawja{} API
presented in Sect.~\ref{sec:api-complete-prog}.  While
\lstinline|invokespecial| and \lstinline|invokestatic| instructions do
not depend on the data of the program, the function used to compute
the result of \lstinline!invokevirtual! and
\lstinline!invokeinterface!  need to be given the set of object types
on which the virtual method may be called.  The analysis needs to have
an over-approximation of the types (classes) of the objects that may
be referenced by the variable on which the method is invoked.

There exists a rich hierarchy of control-flow analyses trading time
for precision~\cite{tip00:call_graph,Grove:Toplas}.
\sawja{} implements the fastest and most cost-effective control-flow
analyses, namely Rapid Type Analysis (RTA) ~\cite{BaconS96},
XTA~\cite{tip00:call_graph} and Class Reachability Analysis (CRA), a
variant of Class Hierarchy Analysis~\cite{dean95:cha}.

\paragraph{Soundness}
\label{sec:soundness}
Our implementation is subject to the usual caveats with respect to
reflection and native methods. As these methods are not written in
\Java{}, their code is not available for analysis and their
control-flow graph cannot be safely abstracted.  Note that our
analyses are always correct for programs that use neither native
methods nor reflection. Moreover, to alleviate the problem, our RTA
implementation can be parametrised by a user-provided abstraction of
native methods specifying the classes it may instantiate and the
methods it may call. A better account of reflection would require an
inter-procedural string analysis~\cite{LivshitsWL05} that is currently
not implemented.

\subsubsection{Implemented Class Analyses}
\label{sec:class-analyses}

\paragraph{RTA}
An object is abstracted by its class and all program variables 
 by the single set of the classes that may
have been instantiated, i.e., this set abstracts all the objects
accessible in the program.  When a virtual call needs to be resolved,
this set is taken as an approximation of the set of objects that may be
referenced by the variable on which the method is called.  This set
grows as the set of reachable methods grows.

\sawja{}'s implementation of RTA is highly optimized.  While static
analyses are often implemented in two steps (a first step in which
constraints are built, and a second step for computing a fixpoint),
here, the program is unknown at the beginning and constraints are
added on-the-fly.  For a faster resolution, we cache all reachable
virtual method calls, the result of their resolution and intermediate
results.  When needed, these caches are updated at every computation
step.
The cached results of method resolutions can then be reused
afterwards, when analyzing the program.

\paragraph{XTA}
As in RTA, an object is abstracted by its class and to every method
and field is attached a set of classes representing the set of object
that may be accessible from the method or field.  An object is
accessible from a method if:
\begin{inparaenum}[(i)]
\item it is accessible from its caller and it is of a sub-type of a
  parameter, or
\item it is accessible from a static field which is read by the
  method,
\item it is accessible from an instance field which is read by the
  method and there an object of a sub-type of the class in which the
  instance fields is declared is already accessible, or
\item it is returned by a method which may be called from the current
  method.
\end{inparaenum}

To facilitate the implementation, we built this analysis on top
of another analysis to refine a previously computed complete program.
This allows us using the aforementioned standard technique (build then
solve constraints).  For the implementation, we need to represent many
class sets. As classes are indexed, these sets can be implemented as
sets of integers.  We need to compute fast union and intersection of
sets and we rarely look for a class in a set.  For those reasons, the
implementation of sets available in the standard library in \OCaml,
based on balanced trees, was not well adapted.  Instead we used a purely 
functional set representation based on Patricia
trees~\cite{morrison68:patricia_trees},  and another
based on BDDs~\cite{bryant92:BDD} (using the external library BuDDy
available at \url{http://buddy.sourceforge.net}).


\paragraph{CRA}
This algorithm computes the complete program without actually
computing the call graph or resolving methods: it considers a
class as \emph{accessible} if it is referenced in another class of the
program, and considers all methods in reachable classes as also
reachable.  When a class references another class, the first one
contains in its constant pool the name of the later one.  Combining
the lazy parsing of our library with the use of the constant pool
allows quickly returning a complete program without even parsing the
content of the methods.
When an actual resolution of method, or a call graph is needed, the
Class Hierarchy Analysis (CHA)~\cite{dean95:cha} is used.
Although parts of the program returned by CRA will be parsed during the
overlying analysis, dead code will never by parsed.



\subsubsection{Experimental Evaluation}

\begin{table}[t]\centering
  \begin{tabular}{|l|l|r|r|r|r|r|r|r|r|}
    \cline{3-10}
    \multicolumn{2}{c|}{} & Soot & Jess & Jml & VNC & ESC/Java & JDTCore & Javacc & JLex\\
    \hline\hline
\multirow{2}{*}{C} 
& CRA & 5,198 & 5,576 & 2,943 & 5,192 & 2,656 & 2,455  & 2,172  & 2,131 \\ \cline{2-10}
& RTA & 4,116 & 2,222 & 1,641 & 1,736 & 1,388  & 1,163  & 792  & 752  \\ \hline\hline
\multirow{5}{*}{M} 
& CRA & 49,810 & 47,122 & 26,906 & 44,678 & 23,229 & 23,579 & 19,389 & 18,485\\ \cline{2-10}
& W-RTA & 32,652 & 4,303  & 17,740 & ? & 9,560 & 7,378 & 3,247 & 1,419\\ \cline{2-10}
& RTA & 32,800 & 12,561 & 11,697 & 9,218 & 8,305 & 9,137 & 4,029 & 3,157 \\ \cline{2-10}
& XTA & 14,251 & 10,043 & 9,408 & 6,534 & 7,039 & 8,186 & 3,250 & 2,392 \\ \cline{2-10}
& W-0CFA & 37,768 & 9,927 & 15,414 & ? & 9,088 & 6,830 & 3,009 & 1,186 \\ \hline\hline
\multirow{5}{*}{E}
& CRA & 2,159,590 & 799,081 & 418,951 & 694,451 & 354,234 & 347,388 & 258,674 & 244,071 \\ \cline{2-10}
& W-RTA & 2,788,533 & 78,444 & 614,216 & ? & 279,232 & 146,119 & 34,192 & 13,256 \\ \cline{2-10}
& RTA & 1,400,958 & 141,910 & 149,209 & 79,029 & 101,257 & 114,454 & 35,727 & 23,209 \\ \cline{2-10}

& XTA & 297,754 & 94,189 & 103,126 & 48,817 & 74,007 & 86,794 & 26,844 & 15,456 \\ \cline{2-10}
& W-0CFA & 856,180 & 183,191 & 187,177 & ? & 87,163 & 77,875 & 21,475 & 4,360 \\ \hline\hline
\multirow{5}{*}{T}
& CRA & 8 & 8 & 4 & 7 & 4 & 5 & 4 & 4\\ \cline{2-10}
& W-RTA & 74 & 7 & 23 & ? & 12 & 12 & 7 & 5 \\ \cline{2-10}
& RTA & 13 & 4 &  4 & 3 &  3 & 4 & 2 & 2\\ \cline{2-10}
& XTA & 187 & 18 & 16 & 11 & 10 & 14 & 5 & 4 \\ \cline{2-10}
& W-0CFA & 2,303 & 209 & 40 & ? & 27 & 26 & 16 & 7 \\ \hline\hline
\multirow{5}{*}{S}
& CRA & 87 & 83 & 51 & 80 & 45 & 47 & 36 & 35 \\ \cline{2-10}
& W-RTA & 248 & 44  & 128 & ? & 84 & 101 & 42 & 8\\ \cline{2-10}
& RTA & 132 & 60 & 54 & 51 & 43 & 52 & 26 & 20 \\ \cline{2-10}
& XTA & 810 & 198 & 184 & 153 & 147 & 157 & 112 & 107 \\ \cline{2-10}
& W-0CFA & 708 & 238 & 215 & ?  & 132 & 134 & 125 & 26\\ \hline\hline
   \end{tabular}
   \caption{Comparison of algorithms generating a program 
     call graph (with \sawja{} and \Wala{}): the different algorithms of 
     \sawja{} (CRA,RTA and XTA) are compared to \Wala{} (W-RTA and W-0CFA)
     with respect to the number of loaded classes (C), reachable methods
     (M) and number of edges (E) in the call graph, their execution
     time (T) in seconds and memory used (S) in megabytes.  Question
     marks (?) indicate clearly invalid results.}
   \label{tab:class-analysis-exp}
\aftertable{}
\end{table}
We evaluate the precision and performances of the class analyses
implemented in \sawja{} on several pieces of \Java{}
software\footnote{Soot (2.3.0), Jess (7.1p1), JML (5.5), TightVNC Java
  Viewer (1.3.9), ESC/Java (2.0b0), Eclipse JDT Core (3.3.0), Javacc
  (4.0) and JLex (1.2.6).} and present our results in
Table~\ref{tab:class-analysis-exp}.
We compared the precision of the 3 algorithms used to compute complete
programs (CRA, RTA and XTA) with respect to the number of reachable
methods in the call graph and its number of edges.  We also give the
number of classes loaded by CRA and RTA.
We provide some results obtained with \Wala{} (version r3767 from the
repository).  Although precision is hard to compare\footnote{Because
  both tools are unsound, a greater number of method in the call graph
  either mean there is a precision loss or that native methods are
  better handled.}, it indicates that, on average, \sawja{} uses half
the memory and time used by \Wala{} per reachable method with RTA.

\section*{Conclusion}
\label{sec:conclusion}

We have presented the \sawja{} library, the first \OCaml{} library
providing state-of-the-art components for writing \Java{} static
analyzers in \OCaml.

The library represents an effort of 1.5 man-year and approximately
22000 lines of \OCaml{} (including comments) of which 4500 are for the
interfaces.  Many design choices are based on our earlier work with
the NIT analyzer~\cite{hubert08-2:nonnull_annotation_inferencer}, a
quite efficient tool, able to analyze a complete program of more than
30000 methods to infer nullness annotations for fields, method
signatures and local variables in less than 2 minutes, while proving
the safety of 80\% of dereferences.  Using our experience from the NIT
development, we designed \sawja{} as a generic framework to allow
every new static analysis prototype to share the same efficient
components as NIT.
Indeed, \sawja{} has already been used in two implementations for the
ANSSI (The French Network and Information Security
Agency)~\cite{jensen10:secure_cloning,hubert10:secure_initialization},
Nit is being ported to the current version of \sawja{}, and other
small analyses (liveness, interval analyses, etc.) are available on
\sawja's web site.
\remarques{Gives the references (link to the archive on their website
  or in a dedicated section of \sawja{}'s website).  Insist on the
  fact that the control flow analyses are already on experiment of the
  lower layers.}

Several extensions are planned for the library. 
Displaying static analysis results is a first challenge that we would
like to tackle. We would like to facilitate the transfer of
annotations from \Java{} source to \Java{} bytecode and then to IR,
and the transfer of analysis results in the opposite direction. We
already provide HTML outputs but ideally the result at source level
would be integrated in an IDE such as Eclipse.  This manipulation has
been already experimented in one of our earlier work for the NIT
static analyzer and we plan to integrate it as a new generic \sawja{}
component.
%
%
To ensure correctness, we would like to replace some components of
\sawja{} by certified extracted code from \Coq{}~\cite{coq}
formalizations.  A challenging candidate would be the IR generation
that relies on optimized algorithms to transform in at most three
passes each bytecode method. We would build such a work on top of the
\textsc{Bicolano}~\cite{bicolano} JVM formalization that has been
developed by some of the authors during the European Mobius project.

\bibliographystyle{plain}
\bibliography{biblio}


\end{document}